\documentclass[sigconf]{acmart}

\renewcommand\footnotetextcopyrightpermission[1]{} 
\setcopyright{none}
\usepackage{csquotes}
\usepackage{caption}
\usepackage{subcaption}

\AtBeginDocument{%
  \providecommand\BibTeX{{%
    \normalfont B\kern-0.5em{\scshape i\kern-0.25em b}\kern-0.8em\TeX}}}

\acmConference[]{}{}{}
\acmBooktitle{}

\newcommand{\app}{iART}

\begin{document}

\title{\app: A Search Engine for Art-Historical Images to Support Research in the Humanities} 



\author{Matthias Springstein}
\affiliation{%
  \institution{\footnotesize TIB -- Leibniz Information Centre for Science and Technology}
  \city{Hanover}
  \country{Germany}
}
\email{matthias.springstein@tib.eu}

\author{Stefanie Schneider}
\affiliation{%
  \institution{\footnotesize Ludwig Maximilian University of Munich}
  \city{Munich}
  \country{Germany}
}
\email{stefanie.schneider@itg.uni-muenchen.de}

\author{Javad Rahnama}
\affiliation{%
  \institution{\footnotesize University Paderborn}
  \city{Paderborn}
  \country{Germany}
}
\email{javad.rahnama@uni-paderborn.de}

\author{Eyke Hüllermeier}
\affiliation{%
  \institution{\footnotesize Ludwig Maximilian University of Munich}
  \city{Munich}
  \country{Germany}
}
\email{eyke@ifi.lmu.de}

\author{Hubertus Kohle}
\affiliation{%
  \institution{\footnotesize Ludwig Maximilian University of Munich}
  \city{Munich}
  \country{Germany}
}
\email{hubertus.kohle@lmu.de}

\author{Ralph Ewerth}
\affiliation{%
  \institution{\footnotesize TIB--Leibniz Information Centre for Science and Technology; \\ 
  L3S Research Center, Leibniz University Hannover}
  \city{Hanover}
  \country{Germany}
}
\email{ralph.ewerth@tib.eu}

\renewcommand{\shortauthors}{Springstein et al.}

\begin{abstract}
In this paper, we introduce \app: an open Web platform for art-historical research that facilitates the process of comparative vision. The system integrates various machine learning techniques for keyword- and content-based image retrieval as well as category formation via clustering. An intuitive GUI supports users to define queries and explore results. By using a state-of-the-art cross-modal deep learning approach, it is possible to search for concepts that were not previously detected by trained classification models. Art-historical objects from large, openly licensed collections such as Amsterdam Rijksmuseum and Wikidata are made available to users.
\end{abstract}





\maketitle

\section{Introduction}


Basic art-historical techniques of analysis are essentially built on comparative processes. Heinrich Wölfflin, e.g., practised comparative vision in determining the stylistic history of the Renaissance and the Baroque, which he interpreted antagonistically \cite{Wolfflin1915}. However, open Web platforms that promote such image-oriented research processes by identifying objects that are similar to each other are currently not available. Previous approaches lack either fine-tuning to the art-historical domain \cite{Rossetto2016}, flexible search query structures that adapt to users' needs \cite{Lang2018}, or the ability for users to upload their own datasets \cite{Offert2021}.

In this paper, we present \app\footnote{\url{https://labs.tib.eu/iart}, accessed: 2021-06-15.}: an e-research-tool that analyses structures (or similarities) of a group of images by processing large, heterogeneous, and digitally available databases of art-historical objects through machine learning. Ordering criteria that were already common in early modern \emph{Wunderkammern}, such as colour, material, or function, can be applied as well as more iconographically based classification principles that, e.g., examine objects for biblical motifs or Christian themes. Decisive is that these principles can be liquefied and reconfigured. The \app\ platform has been designed in such a way that it can be easily extended with the help of plug-ins to meet the various requirements of art historians. The retrieval of objects is not only performed with automatically generated keywords, but also by utilizing state-of-the-art multimodal embeddings that enable search based on accurate, detailed scene descriptions given by the user.

\begin{figure*}
    \centering
    \begin{subfigure}[b]{0.38\linewidth}
        \centering
        \includegraphics[width=\linewidth]{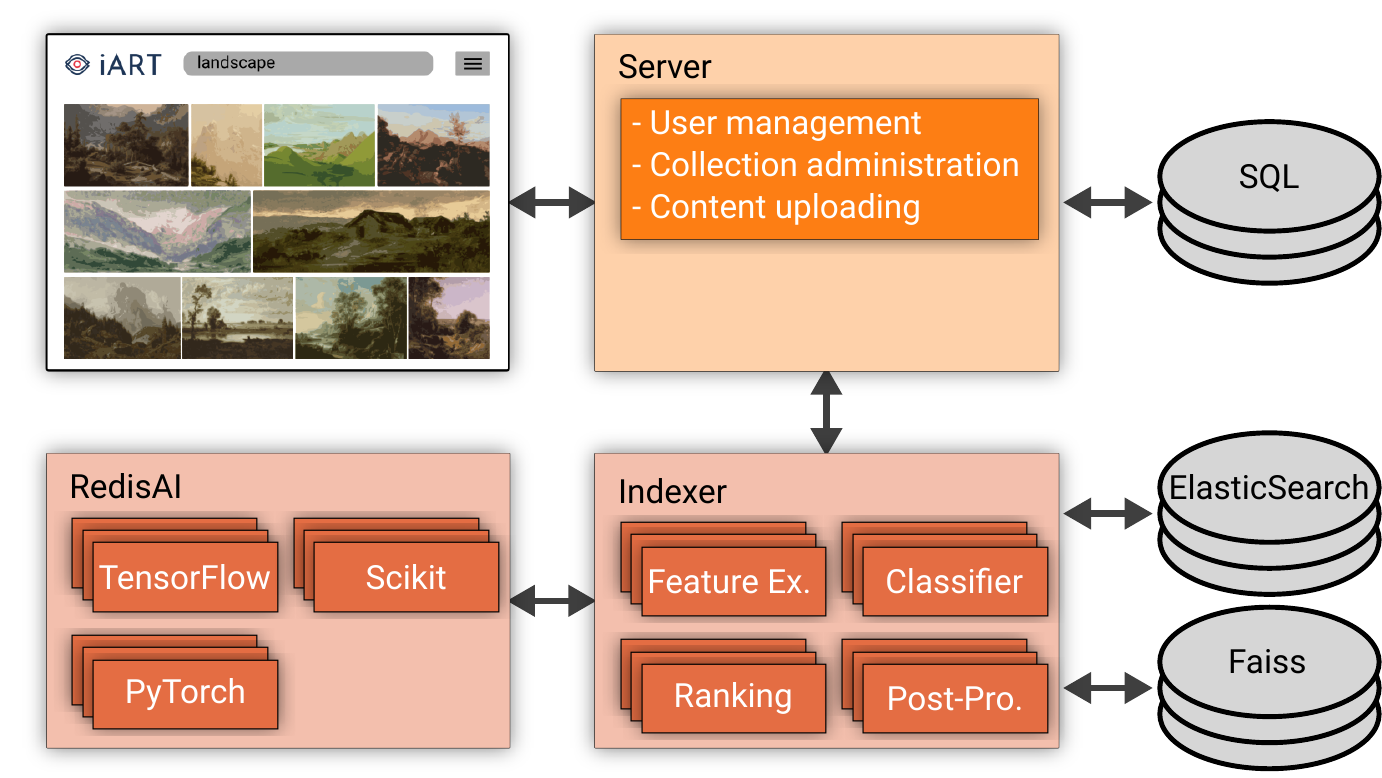}
        \caption{}
        \label{fig:arch}
    \end{subfigure}
    \hfill
    \begin{subfigure}[b]{0.58\linewidth}
        \centering
        \raisebox{5mm}{\includegraphics[width=\textwidth]{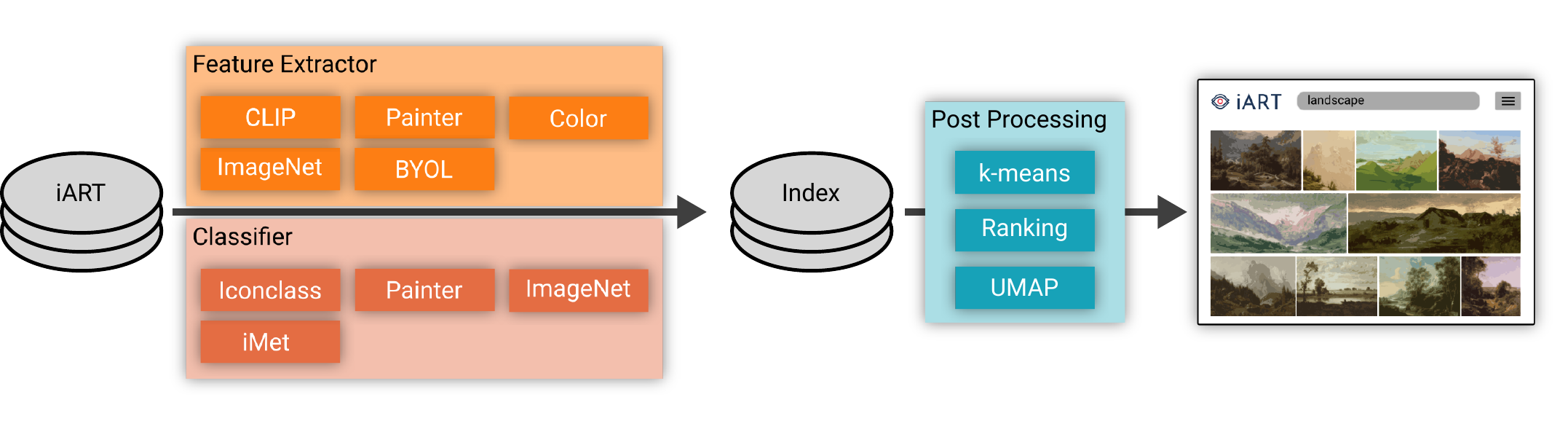}}
        \caption{}
        \label{fig:pipeline}
    \end{subfigure}
    
    \caption{(a) Architecture with associated database structure and RedisAI inference server; (b) indexing and post-processing steps with relevant plug-ins. The generated keywords and features are stored with the help of Elasticsearch and Faiss \cite{DBLP:journals/corr/JohnsonDJ17}.}
    \label{fig:iart}
\end{figure*}

\section{System Architecture}
To facilitate adaptation to diverse research interests, the software is designed to be as modular as possible: the individual indexing steps are outsourced to plug-ins and user administration is separated from the search infrastructure. All models are accelerated with a RedisAI inference server to optimally manage the resources needed for computation. This step makes it easier to run different deep learning models on a single GPU and enables the use of back end systems such as PyTorch or TensorFlow. The Web front end integrates the Vue framework and communicates with the indexer via a Django Web service. Figure~\ref{fig:arch} displays the architecture of \app.

\section{Image Retrieval}
Due to the diffuseness of the concept of similarity, \app\ supports different types of search queries that can be precisely targeted by the user. Hence, the underlying system was developed with the idea that each processing step should be extensible through a plug-in structure. This applies in particular to feature extraction, image classification, ranking of results, and various post-processing steps that serve visualisation and clustering. The complete pipeline is shown in Figure~\ref{fig:pipeline}.

Common feature extractors are integrated in \app, e.g., for ImageNet embeddings through a pre-trained ResNet \cite{DBLP:conf/cvpr/HeZRS16}. These are complemented by models adjusted to the art-historical domain: (1) the self-supervised model BYOL (Bootstrap Your Own Latent) is trained on an adequate subset of Wikimedia images~\cite{DBLP:conf/nips/GrillSATRBDPGAP20}; (2) while the Painter model utilizes the Painter by Numbers dataset~\cite{painter} to extract features for style and genre; (3) and the transformer-based neural network CLIP (Contrastive Language-Image Pre-Training) learns visual concepts from natural language supervision~\cite{DBLP:journals/corr/abs-2103-00020}. Moreover, different classification models are trained to automatically predict art-historically relevant phenomena collected from Iconclass~\cite{iconclass}, iMet~\cite{imet}, and Painter by Numbers~\cite{painter}. 

The extracted features primarily enable the user to retrieve similar images based on a query image. As the system extracts different embeddings for each image, the user can change the weighting of plug-ins and thus adjust the order of results according to his or her needs. Through its two decoder structures, CLIP creates a unified feature space for image and text, allowing the user to also enter textual descriptions. To create more complex queries, multiple reference documents with different weights can be processed. For example, a reference image of Saint Sebastian can be combined with the text query \enquote{crucifixion} (Figure~\ref{fig:sebastian-ranked}). Using a faceted search, the list of results can be further narrowed down based on classified attributes and the metadata given by the respective collection. This helps users to filter their uploaded inventories and 
about one million openly licensed images, including examples from Amsterdam Rijksmuseum~\cite{rijksmuseum}, Wikidata~\cite{wikidata}, Kenom~\cite{kenom}, and ARTigo~\cite{Wieser2013}\cite{Becker2018}.

\section{Result Visualization}
\begin{figure}
    \centering
    \begin{subfigure}[b]{0.5\textwidth}
        \centering
        \includegraphics[width=\textwidth]{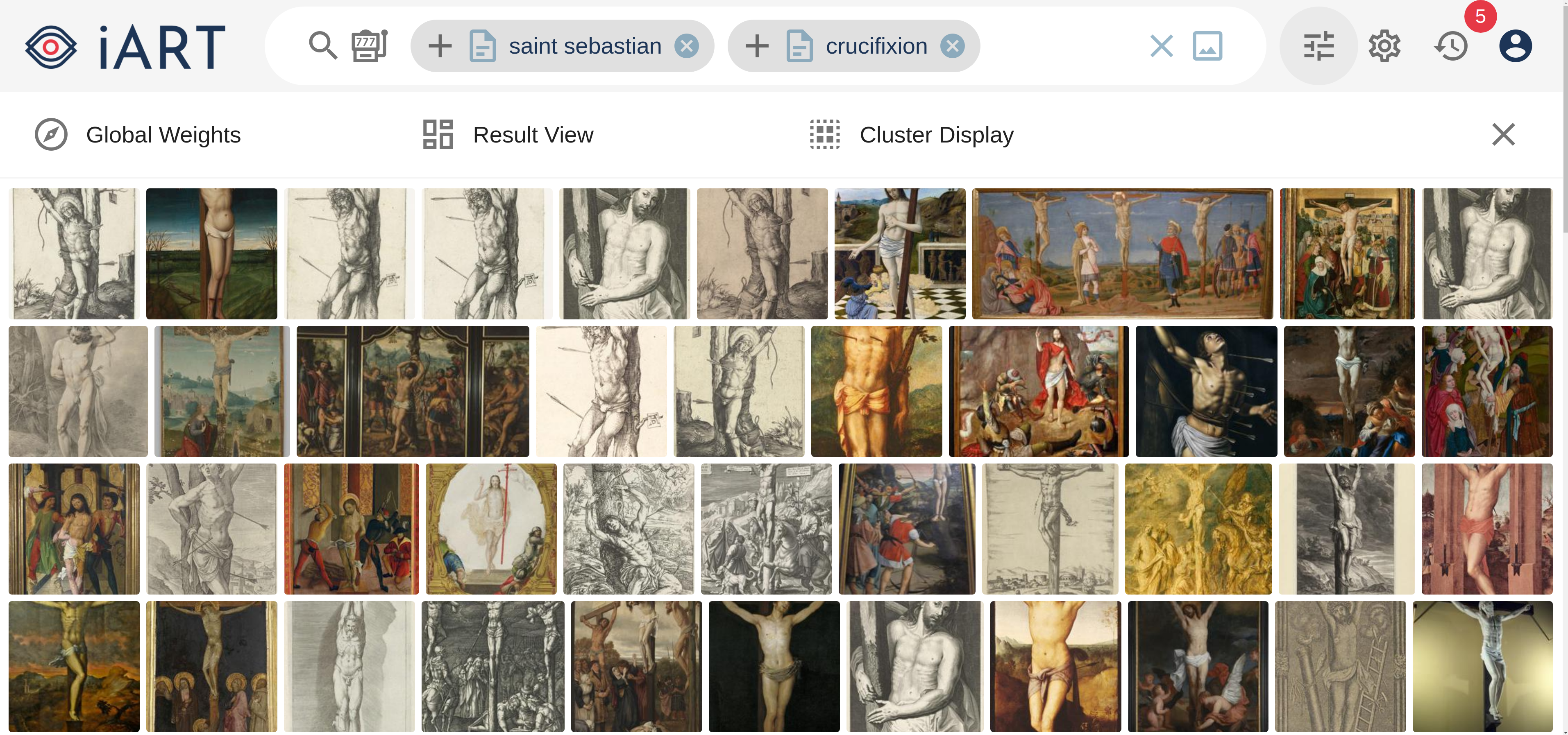}
        \caption{}
        \label{fig:sebastian-ranked}
    \end{subfigure}
    \hfill
    \begin{subfigure}[b]{0.5\textwidth}
        \centering
        \includegraphics[width=\textwidth]{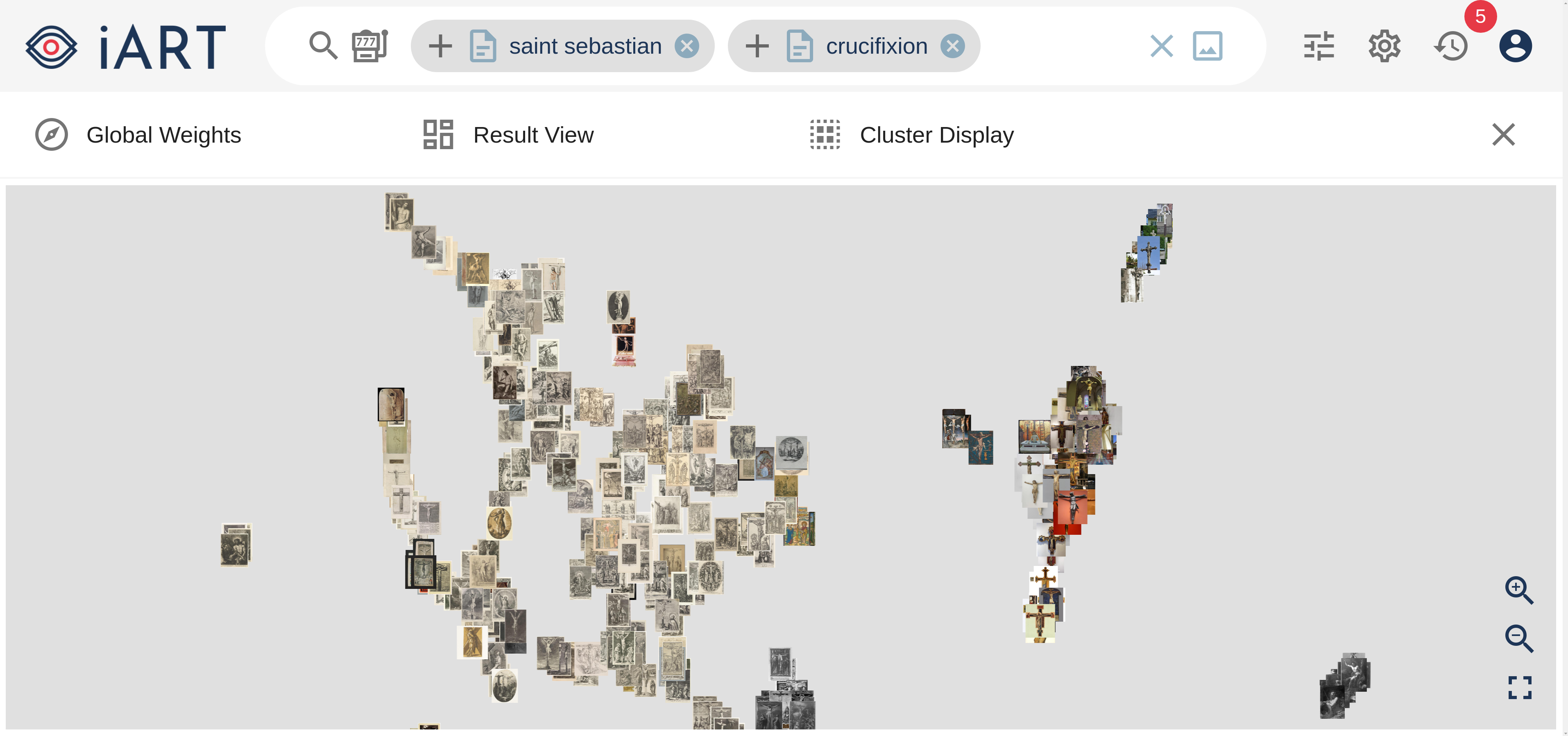}
        \caption{}
        \label{fig:sebastian-umap}
    \end{subfigure}
    
    \caption{Search results for a reference image of Saint Sebastian, combined with the text query \enquote{crucifixion.} (a) Default image grid; (b) two-dimensional canvas view.}
    \label{fig:sebastian}
\end{figure}

To simplify the exploration of the results, different object views are implemented. By default, an image grid sorted by relevance is displayed, via which further details are provided on demand, such as metadata from the respective object. Results can be clustered with k-means and visualized, e.g., as image carousels vertically separated by groups. For more advanced use cases, it is possible to arrange the images on a two-dimensional canvas using the dimensionality reduction technique UMAP (Uniform Manifold Approximation and Projection)~\cite{DBLP:journals/corr/abs-1802-03426}. Zoom and filter operations, such as an interactive drag-select to juxtapose multiple objects, are supported with the aid of VisJs (Figure~\ref{fig:sebastian-umap}).


\section{Conclusion}

With \app, we introduced an open Web platform for art-historical research that facilitates the process of comparative vision. As the system is extensible and supports various classification plug-ins and feature extractors, users can adapt it to their needs. In the future, the system will be enriched with additional openly licensed datasets. Further plug-ins will also be integrated, e.g., feature extractors for human body poses or image composition.


\bibliographystyle{ACM-Reference-Format}
\bibliography{references}


\end{document}